\documentstyle[11pt]{article}

\newcommand{\B}{{\cal B}}

\begin{document}
\begin{center}
{\LARGE{\bf Energy and Momentum densities of  cosmological models,
with equation of state $\rho=\mu$, in general relativity and
teleparallel
gravity }}\\[2em]
\large{\bf{Ragab M. Gad}\footnote{Email Address: ragab2gad@hotmail.com}}\\
\normalsize {Mathematics Department, Faculty of Science,}\\
\normalsize  {Minia University, 61915 El-Minia,  EGYPT.}
\end{center}

\begin{abstract}
We calculated  the energy and momentum densities of stiff fluid
solutions, using Einstein,  Bergmann-Thomson and Landau-Lifshitz
energy-momentum complexes, in both general relativity and
teleparallel gravity.  In our analysis we get different results
comparing  the aforementioned complexes with each other when
calculated in the same gravitational theory, either this is in
general relativity and teleparallel gravity. However,
interestingly enough, each complex's value is the same either in
general relativity or teleparallel gravity. Our results sustain
that (i) general relativity or teleparallel gravity   are
equivalent theories (ii) different energy-momentum complexes do
not provide the same energy and momentum densities neither in
general relativity nor in teleparallel gravity. In the context of
the theory of teleparallel gravity, the vector and axial-vector
parts of the torsion are obtained. We show that the axial-vector
torsion vanishes for the space-time under study.

\end{abstract}

\setcounter{equation}{0}
\section{Introduction}
The issue of energy localization was first discussed during the
early years after the development of general relativity and debate
continued for decades.  There are different attempts to find a
general accepted definition of the energy density for the
gravitational field. However, there is still no generally accepted
definition known. The foremost endeavor was made by Einstein
\cite{E} who suggested a definition for energy-momentum
distribution. Following this definition, many physicists proposed
different energy-momentum complexes: e.g. Tolman  \cite{T}, Landau
and Lifshitz  \cite{LL}, Papapetrou \cite{P}, Bergmann and Thomson
\cite{B}, Weinberg  \cite{W} and M{\o}ller \cite{E1}. Except for
the M{\o}ller definition, others are restricted to calculate the
energy and momentum distributions in quasi-Cartesian coordinates
to get a reasonable and meaningful result.
\par
Despite these drawbacks, some interesting results obtained
recently leads to the conclusion that these definitions give
exactly the same energy distribution for any given space-time
\cite{V1}-\cite{V97}. However, some examples of space-times have
been explored which do not support these results
\cite{Gad1}-\cite{g2}.
\par
The problem of energy-momentum localization can also be
reformulated in the context of teleparallel gravity \cite{24}. By
working in the context of teleparallel gravity, Vargas \cite{24}
obtained  the teleparallel version of both Einstein and
Landau-Lifshitz energy-momentum complexes. He used these
definitions and found that the total energy is zero in
Friedmann-Robertson-Walker space-time. His results are the same as
those calculated in general relativity. Salti and his
collaborators \cite{S}-\cite{HKS} considered different space-times
for various definitions in teleparallel gravity to obtain the
energy-momentum distribution in a given model. Their results agree
with the previous results obtained in the theory of general
relativity.
\par

The paper is organized as follows: In the next section we briefly
present the cosmological model, whose source is a stiff fluid, in
which the energy and momentum densities by using different
prescriptions are to be calculated. In three subsections of
section 3, we explicitly compute the energy and momentum densities
of the space-time under consideration, using Einstein,
Bergmann-Thomson and Landau-Lifshitz energy momentum complexes, in
the context of general relativity. In section 4, we briefly
present the concept of energy-momentum complexes in the context of
teleparallel gravity theory. In the subsequent section, we
calculate the energy and momentum densities in teleparallel
gravity, using the aforementioned complexes. In section 6, we
obtain the vector and axial-vector parts of the torsion in the
theory of teleparallel gravity. Finally, in section 7, a brief
summary of results and concluding remarks are presented.
\par
Through this paper we will use $G = 1$ and $c = 1$ units and  the
Greek alphabet $(\mu, \nu , \rho, · · · = 0, 1, 2, 3)$ to denote
tensor indices, that is, indices related to space-time. The Latin
alphabet $(a, b, c, · · · = 0, 1, 2, 3)$ will be used to denote
local Lorentz (or tangent space) indices, whose associated metric
tensor is $\eta_{ab}= diag(-1,1,1,1)$. \setcounter{equation}{0}
\section{Stiff Fluid Solutions}
The cylindrically symmetric solutions of the Einstein equations
whose source is a "stiff fluid", that is, a perfect fluid with the
equation of state, $\rho = p$, can be written in the form
\begin{equation}\label{2}
ds^2 = g_{00}(-dt^2 + dx^2)+ g_{AB}dx^Adx^B,
\end{equation}
where $A, B=2,3$, $(x^2, x^3)=(y,z)$ and all the components
$g_{\alpha\beta}$ depend only on $t$ and $x$.
\par
The interest in vacuum solutions with the above metric stems from
the fact that there exists an elaborate theory which tells how to
generate new vacuum solutions  of the form (\ref{2}) from known
solutions by algebraic operations (the technique and its
applications were given by Verdaguer \cite{Ver}).
\par
The possibility of generating cylindrically symmetric "stiff
fluid" solutions from vacuum solutions was apparently first
indicated by Wainwright et al \cite{Wainw}. These solutions have
all been generated by an algorithm.
\par
The metric discussed by Wainwright et al \cite{Wainw}  is of the
form
\begin{equation}\label{1}
ds^2 = e^{2k+\Omega}( - dt^2 + dx^2) + R[f(dy+\omega dz)^2
+f^{-1}dz^2],
\end{equation}
where $k, \Omega, R, f$ and $\omega$ are functions of $t$ and
$x$.\\
This metric admits the two commuting spacelike Killing vector
fields \footnote{A vector field $\xi$ which satisfied
$\pounds_{\xi} g_{\mu\nu} =\xi_{\mu;\nu} + \xi_{\nu;\mu} =0$ is
called Killing vector field.}
$$
\frac{\partial}{\partial y}, \qquad \frac{\partial}{\partial z}.
$$
Thus it must satisfy the field equation
$$
{\bf{R}}_{\mu\nu}-\frac{1}{2}g_{\mu\nu}{\bf{R}}=8\pi
[(\rho+p)u_{\mu}u_{\nu} + pg_{\mu\nu}].
$$
With
$$
\rho =p.
$$
The covariant and contravariant components of the metric
(\ref{1}), respectively, are given as follows
\begin{equation}\label{2-4}
g_{\mu\nu} =-e^{2k+\Omega}\delta^{0}_{\mu}\delta^{0}_{\nu}+
e^{2k+\Omega}\delta^{1}_{\mu}\delta^{1}_{\nu}+
Rf\delta^{2}_{\mu}\delta^{2}_{\nu}+ (Rf\omega^2
+f^{-1})\delta^{3}_{\mu}\delta^{3}_{\nu}+
Rf\omega(\delta^{2}_{\mu}\delta^{3}_{\nu}+
\delta^{3}_{\mu}\delta^{2}_{\nu}),
\end{equation}
\begin{equation}\label{2.4}
g^{\mu\nu} =-e^{-(2k+\Omega)}\delta_{0}^{\mu}\delta_{0}^{\nu}+
e^{-(2k+\Omega)}\delta_{1}^{\mu}\delta_{1}^{\nu}+
\frac{f^2\omega^2+1}{Rf}\delta_{2}^{\mu}\delta_{2}^{\nu}+
\frac{f}{R}\delta_{3}^{\mu}\delta_{3}^{\nu}
-\frac{f\omega}{R}(\delta_{2}^{\mu}\delta_{3}^{\nu}+
\delta_{3}^{\mu}\delta_{2}^{\nu}).
\end{equation}
 \setcounter{equation}{0}
\section{Energy-momentum complexes in general relativity}
\subsection{ Einstein's Energy-momentum Complex}
The energy-momentum complex as defined by Einstein \cite{E} is
given by
\begin{equation} \label{3.1}
\theta^{\nu}_{\mu} = \frac{1}{16\pi}H^{\nu\alpha}_{\, \,\,
\mu,\alpha},
\end{equation}
where the Einstein's superpotential $H^{\nu\alpha}_{\, \,\, \mu}$
is of the form
\begin{equation} \label{3.2}
H^{\nu\alpha}_{\, \,\, \mu} = - H^{\alpha\nu}_{\, \,\, \mu} =
\frac{g_{\mu\beta}}{\sqrt{- g}} \big[ - g\big(
g^{\nu\beta}g^{\alpha\rho} -
g^{\alpha\beta}g^{\nu\rho}\big)\big]_{,\rho}.
\end{equation}
$\theta^{0}_{0}$ and $\theta^{0}_{i}, (i=1,2,3),$ are the energy
and
momentum density components, respectively.\\
The energy-momentum complex $\theta^{\nu}_{\mu}$ satisfies the
local conservation law
$$
\frac{\partial\theta^\nu_\mu}{\partial x^\nu} = 0
$$
\par
In order to evaluate the energy and momentum densities in
Einstein's prescription associated with the space-time under
consideration, we evaluate the non-zero components of
$H^{\nu\alpha}_{\, \,\, \mu}$

\begin{equation}\label{3.5}
\begin{array}{ccc}
H^{01}_{\, \,\, 0}& =2R^{\prime},
\\
H^{01}_{\, \,\, 1} &= 2\dot{R}
\\
 H^{02}_{\, \,\, 2} & =f\big(\frac{f\dot{R}-R\dot{f}}{f^2}
 +\frac{R}{f}(2\dot{k}+\dot{\Omega})+rf\omega\dot{\omega}\big),\\
H^{02}_{\, \,\, 3} &= -Rf^2\dot{\omega},\\
H^{03}_{\, \,\, 3} &= f\omega\big( -Rf\dot{\omega}
+\frac{R}{f\omega}(2k+\dot{\Omega}) + \frac{R\dot{f}}{\omega f^2}
+ \frac{\dot{R}}{f\omega}\big).
\end{array}
\end{equation}
Using these components  in equation (\ref{3.1}), we get the energy
and momentum densities as following
$$
\begin{array}{ccc}
 \theta^{0}_{0}& =\frac{1}{8\pi}R^{\prime\prime},\\
\theta^{0}_{1}& =\frac{1}{8\pi}\dot{R}^\prime,\\
\theta^{0}_{2} &=\theta^{0}_{3} =0.
\end{array}
$$
\subsection{The Energy-Momentum
Complex of Bergmann-Thomson}
The Bergmann-Thomson energy-momentum
complex \cite{B} is given by
\begin{equation}\label{6.1}
B^{\mu\nu} =
\frac{1}{16\pi}\big[g^{\mu\alpha}\B^{\nu\beta}_{\alpha}\big]_{,\beta},
\end{equation}
where
$$
\B^{\nu\beta}_{\alpha} =
\frac{g_{\alpha\rho}}{\sqrt{-g}}\Big[-g\Big(g^{\nu\rho}g^{\beta\sigma}-
g^{\beta\rho}g^{\nu\sigma}\Big)\Big]_{,\sigma}.
$$
$B^{00}$ and $B^{0i}, (i=1,2,3),$ are the energy and momentum
density components. In order to calculate $B^{00}$ and $B^{0i}$
for Weyl metric, using Bergmann-Thomson  energy-momentum complex,
we require the following non-vanishing components of
$\B^{\nu\beta}_{\alpha}$
\begin{equation}\label{6.5}
\begin{array}{ccc}
\B^{01}_{\, \,\, 0}& =2R^{\prime},
\\
\B^{01}_{\, \,\, 1} &= 2\dot{R}
\\
 \B^{02}_{\, \,\, 2} & =f\big(\frac{f\dot{R}-R\dot{f}}{f^2}
 +\frac{R}{f}(2\dot{k}+\dot{\Omega})+rf\omega\dot{\omega}\big),\\
\B^{02}_{\, \,\, 3} &= -Rf^2\dot{\omega},\\
\B^{03}_{\, \,\, 3} &= f\omega\big(
-Rf\dot{\omega}+\frac{R}{f\omega}(2k+\dot{\Omega}) +
\frac{R\dot{f}}{\omega f^2} + \frac{\dot{R}}{f\omega}\big).
\end{array}
\end{equation}

 Using the components (\ref{6.5}) and (\ref{2.4}) in (\ref{6.1}), we get
the energy and momentum densities for the space-time under
consideration, respectively, as follows
\begin{equation}
\begin{array}{ccc}
B^{00}&
=\frac{e^{-(2k+\Omega)}}{8\pi}\big[(2k^{\prime}+\Omega^{\prime})R^{\prime}-R^{\prime\prime}\big],
\\
B^{01}
&=\frac{e^{-(2k+\Omega)}}{8\pi}\big[\dot{R}^{\prime}-(2\dot{k}+\dot{\Omega})R^{\prime}\big],
\\
 B^{02}& = B^{03}=0.
\end{array}
\end{equation}

\subsection{ Landau-Lifshitz's Energy-momentum Complex}
The energy-momentum complex of Landau-Lifshitz \cite{LL} is
\begin{equation}\label{ll1}
L^{\mu\nu}= \frac{1}{16\pi}\S^{\mu\alpha\nu\beta}_{\quad
,\alpha\beta},
\end{equation}
where $\S^{\mu\alpha\nu\beta}$ with symmetries of the Riemann
tensor and is defined by
\begin{equation}\label{LL2}
\S^{\mu\alpha\nu\beta}=-g(g^{\mu\nu}g^{\alpha\beta}-g^{\mu\beta}g^{\alpha\nu}).
\end{equation}
The quantity $L^{00}$ represents the energy density of the whole
physical system including gravitation and $L^{0i}, (i=1,2,3),$
represents the components of the total momentum (energy current)
density.
\par
In order to evaluate the energy and momentum densities in
Landau-Lifshitz's prescription associated with the  metric
(\ref{1}), we evaluate the non-zero components of
$\S^{\mu\alpha\nu\beta}$

\begin{equation}\label{3.5}
\begin{array}{ccc}
\S^{0101} & =-R^2 ,\\
\S^{0202} & =-\frac{R}{f}(f^2\omega^2 +1)e^{2k+\Omega} ,\\
\S^{0203} & = f\omega Re^{2k+\Omega},\\
\S^{0303} & = -Rfe^{2k+\Omega}.
\end{array}
\end{equation}
Using these components  in equation (\ref{ll1}),  we get the
energy and momentum densities as following
\begin{equation}
\begin{array}{ccc}
 L^{00}& =-\frac{1}{8\pi}(RR^{\prime\prime}+R^{\prime 2}),\\
L^{01}& =\frac{1}{8\pi}[R\dot{R}^{\prime}+\dot{R}R^{\prime}],\\
L^{02}& = L^{03}=0.
\end{array}
\end{equation}
\setcounter{equation}{0}
\section{Energy-momentum Complexes in Teleparallel Gravity}
The name of teleparallel gravity is normally used to denote the
general three-parameter theory introduced in \cite{HS}. The
teleparallel gravity equivalent of general relativity \cite{H02}
can indeed be understand as a gauge theory for the translation
group based on Weitzenb\"{o}ck geometry \cite{Wei}. In this
theory, the gravitational interaction is described by a force
similar to Lorentz force equation of electrodynamics, with torsion
playing the role of force  \cite{deA} and the curvature tensor
vanishes identically.
\par
Let us start by reviewing the fundamentals of the teleparallel
equivalent of general relativity (see for example
\cite{HS,deA,deAG}. A gauge transformations is defined as a local
translation of the tangent coordinates,
$$
x^a \rightarrow x^{\prime a} = x^a + \eta^a,
$$
where $\eta^a(x^{\nu})$ are the transformation parameter. For an
infinitesimal transformation, we have
$$
\delta x^a = \delta\eta^c P_c x^a
$$
with $P_{c}=\frac{\partial}{\partial x}$ the generators of
transformation. The gauge covariant derivative of a general matter
field $\Phi (x^{\nu} )$ is
$$
D_{\nu}\Phi =h^{a}_{\quad \nu} \partial_{a}\Phi,
$$
where
\begin{equation} \label{pot}
h^{a}_{\,\, \nu} =\partial_{\nu} x^a + B^{a}_{\, \, \nu}
\end{equation}
is a non-trivial tetrad field, $B^{a}_{\,\, \nu}$ is the
translation gauge potential. \\
The relation (\ref{pot}) satisfies the orthogonality condition
\begin{equation}\label{orth}
h^{a}_{\,\, \nu}h_{a}^{\,\, \nu}= \delta_{\mu}^{\nu}.
\end{equation}
Notice that, where as the tangent space indices are raised and
lowered with the Minkowski metric $\eta_{ab}$, the space-time
indices are raised and lowered with the space-time metric
\begin{equation} \label{}
g_{\mu\nu} =\eta_{ab} h^{a}_{\,\, \mu} h^{b}_{\,\, \nu}.
\end{equation}
A nontrivial tetrad field induces on space-time a teleparallel
structure which is directly related to the presence of the
gravitational field. The parallel transport of the tetrad
$h^{a}_{\,\, \nu}$ between two neighboring points is encoded in
the covariant derivative
$$
\nabla_{\mu} h^{a}_{\,\, \nu} =\partial h^{a}_{\,\, \nu} -
\Gamma^{\alpha}_{\mu\nu}h^{a}_{\,\, \alpha},
$$
where
\begin{equation}\label{conn}
\Gamma^{\alpha}_{\,\,\mu\nu}= h_{a}^{\,\, \alpha}\partial_{\mu}
h^{a}_{\,\, \nu}
\end{equation}
is the Weitzenb\"{o}ck connection. This connection is presenting
torsion, but no curvature. The torsion of the Weitzenb\"{o}ck
connection is defined by
\begin{equation}\label{tor}
T^{\rho}_{\,\,\mu\nu}=\Gamma^{\rho}_{\,\,\nu\mu} -
\Gamma^{\rho}_{\,\,\mu\nu}.
\end{equation}
The Lagrangian of the teleparallel  equivalent of general
relativity is given by
$$
\pounds =\pounds_{G} +\pounds_{M} = \frac{c^4h}{16\pi
G}S^{\rho\mu\nu}T_{\rho\mu\nu} +\pounds_{M},
$$
where $h=det(h^a_{\,\,\mu}$, $\pounds_{M}$ is the Lagrangain of a
source field and
\begin{equation}\label{S}
S^{\rho\mu\nu}= c_{1}T^{\rho\mu\nu} +
\frac{c_{2}}{2}\big(T^{\mu\rho\nu} - T^{\nu\rho\mu}\big) +
\frac{c_{3}}{2}\big(g^{\rho\nu}T^{\sigma\mu}_{\,\,\,\,\sigma} -
g^{\mu\rho}T^{\sigma\nu}_{\,\,\,\,\sigma}\big).
\end{equation}
is a tensor written in terms of Weitzenb\"{o}ck connection. In the
above form $c_{1}, c_{2}$ and $c_{3}$ are the three
dimensionalless coupling constants of teleparallel gravity.\\
For the so called teleparallel equivalent of general relativity,
the specific choice of these constants are given by \cite{HS}
\begin{equation}
c_{1} =\frac{1}{4}, \qquad c_{2}=\frac{1}{2}, \qquad c_{3}=-1.
\end{equation}
The energy-momentum complexes of Einstein, Bergmann-Thomson and
Landau-Lifshitz in teleparallel gravity, respectively, are given
by \cite{24}
\begin{equation}\label{EBL}
\begin{array}{ccc}
hE^{\mu}_{\,\,\,\,\nu} & = \frac{1}{4\pi}\partial_{\lambda}\Big(\mho_{\nu}^{\,\,\mu\lambda}
\Big),\\
hB^{\mu\nu} & = \frac{1}{4\pi}\partial_{\lambda}\Big(g^{\mu\beta}\mho_{\beta}^{\,\,\nu\lambda}\Big),\\
hL^{\mu\nu} & =
\frac{1}{4\pi}\partial_{\lambda}\Big(hg^{\mu\beta}\mho_{\beta}^{\,\,\nu\lambda}\Big),
\end{array}
\end{equation}
where $\mho_{\nu}^{\,\,\mu\lambda}$ is the Freud's super-potential
and defined as follows
\begin{equation}\label{U}
\mho_{\nu}^{\,\,\mu\lambda}=hS_{\nu}^{\,\,\mu\lambda}
\end{equation}
 The energy and momentum distributions in
the above complexes, respectively, are
\begin{equation}\label{e-mD}
\begin{array}{ccc}
P^{E}_\mu & = \int_{\Sigma}hE^0_{\,\,\mu}d^3x,\\
P^{BT}_\mu & = \int_{\Sigma}hB^0_{\,\,\mu}d^3x,\\
P^{LL}_\mu & = \int_{\Sigma}hL^0_{\,\,\mu}d^3x,
\end{array}
\end{equation}
where $P_{0}$ is the energy, $P_{i}\quad (i=1,2,3)$ are the
momentum components and the integration hypersurface $\Sigma$ is
described by $x^0 =t$ constant.
\setcounter{equation}{0}
\section{Energy and momentum associated with the metric (\ref{1})
in Teleparallel gravity} For the line element (\ref{1}), using
(\ref{}), we obtain the tetrad   components
\begin{equation}\label{tetrad}
\begin{array}{ccc}
h^0_{\,\, 0} & = e^{\frac{1}{2}(2k+\Omega)},\\
h^1_{\,\, 1} & = e^{\frac{1}{2}(2k+\Omega)},\\
h^2_{\,\, 2} & = \sqrt{Rf},\\
h^2_{\,\, 3} & = \omega\sqrt{Rf},\\
h^3_{\,\, 3} & = \sqrt{\frac{R}{f}},
\end{array}
\end{equation}
 its inverse
\begin{equation}\label{tetrad1}
\begin{array}{ccc}
h_0^{\,\, 0} & = e^{-\frac{1}{2}(2k+\Omega)},\\
h_1^{\,\, 1} & = e^{-\frac{1}{2}(2k+\Omega)},\\
h_2^{\,\, 2} & = \frac{1}{\sqrt{Rf}},\\
h_3^{\,\, 2} & = -\omega\sqrt{\frac{R}{f}},\\
h_3^{\,\, 3} & = \sqrt{\frac{f}{R}}.
\end{array}
\end{equation}
\begin{equation}
h=det(h^a_{\,\,\mu})=Re^{(2k+\Omega)}.
\end{equation}
 Using the above tetrad and its inverse in
equation (\ref{conn}), we get the following non-vanishing
Weitzenb\"{o}ck connection components
\begin{equation}\label{gammas}
\begin{array}{ccc}
\Gamma^{0}_{\,\, 00} & = \frac{1}{2}(2\dot{k}+\dot{\Omega}),\\
\Gamma^{0}_{\,\, 01} & = \frac{1}{2}(2k^{\prime}+\Omega^{\prime}),\\
\Gamma^{1}_{\,\, 10} & = \frac{1}{2}(2\dot{k}+\dot{\Omega}),\\
\Gamma^{1}_{\,\, 11} & = \frac{1}{2}(2k^{\prime}+\Omega^{\prime}),\\
\Gamma^{2}_{\,\, 20} & =\frac{R\dot{f} +f\dot{R}}{2Rf} ,\\
\Gamma^{2}_{\,\, 21} & =\frac{Rf^{\prime} +fR^{\prime}}{2Rf} ,\\
\Gamma^{2}_{\,\, 30} & =\frac{\omega\dot{f}}{f} ,\\
\Gamma^{2}_{\,\, 31} & =\frac{\omega f^{\prime}}{f} ,\\
\Gamma^{3}_{\,\,30} & =\frac{R\dot{f} - f\dot{R}}{2R^2} ,\\
\Gamma^{3}_{\,\, 31} & =\frac{Rf^{\prime} +fR^{\prime}}{2R^2},
\end{array}
\end{equation}
where dot and prime indicates derivative with respect to $t$ and
$x$, respectively. Using (\ref{tor}) and the above components, we
find the following non-vanishing torsion components
\begin{equation}\label{torsion}
\begin{array}{ccc}
T^{0}_{\,\, 10} & = - T^{0}_{\,\, 01} & = \frac{1}{2}(2k^{\prime}+\Omega^{\prime}),\\
T^{1}_{\,\, 01} & = - T^{1}_{\,\, 10} & = \frac{1}{2}(2\dot{k}+\dot{\Omega}),\\
T^{2}_{\,\, 02} & = -T^{2}_{\,\, 20} & = \frac{R\dot{f} +f\dot{R}}{2Rf} ,\\
T^{2}_{\,\, 12} & = -T^{2}_{\,\, 21} & = \frac{Rf^{\prime} +fR^{\prime}}{2Rf} ,\\
T^{2}_{\,\, 03} & = -T^{2}_{\,\, 30} & = \frac{\omega\dot{f}}{f} ,\\
T^{2}_{\,\, 13} & = - T^{2}_{\,\, 31} & = \frac{\omega f^{\prime}}{f} ,\\
T^{3}_{\,\, 03} & = - T^{3}_{\,\, 30} & = \frac{f\dot{R} - R\dot{f}}{2Rf} ,\\
T^{3}_{\,\, 13} & = - T^{3}_{\,\, 31} & = \frac{fR^{\prime}
-Rf^{\prime}}{2Rf}.
\end{array}
\end{equation}
Using these results into equation (\ref{S}), the non-vanishing
components of the tensor $S^{\,\,\mu\nu}_{\beta}$ are as following
\begin{equation}\label{Scalc}
\begin{array}{ccc}
S_{0}^{\,\, 01} & = \frac{R^\prime}{2R}e^{-(2k+\Omega)},\\
S_{1}^{\,\, 01} & = \frac{\dot{R}}{2R}e^{-(2k+\Omega)},\\
S_{2}^{\,\, 02} & =  e^{-(2k+\Omega)}\Big[ \frac{3\dot{f}}{f}-
\frac{3}{4}f\dot{f}\omega^2 - \frac{\dot{R}}{2R} -\dot{k}-\frac{\dot{\Omega}}{2}\Big],\\
S_{3}^{\,\, 02} & = -\frac{1}{4}e^{-(2k+\Omega)}\dot{f}\omega\big[\omega^2 +1 +2f^{-1}\big] ,\\
S_{3}^{\,\, 03} & =  e^{-(2k+\Omega)}\Big[
\frac{3\dot{R}}{4R}+\frac{1}{4}f\dot{f}\omega^2
+\dot{k}+\frac{\dot{\omega}}{2}\Big].
\end{array}
\end{equation}
Using these components and the relation (\ref{U}) in equation
(\ref{EBL}), we obtain the energy and momentum densities in the
sense of Einstein, Bergmann-Thomson and Landau-Lifshitz,
respectively, as follows
\begin{equation}
\begin{array}{ccc}
hE^{0}_{0}& =\frac{1}{8\pi}R^{\prime\prime},\\
hE^{0}_{1}& =\frac{1}{8\pi}\dot{R}^{\prime},\\
hE^{0}_{2} &=hE^{0}_{3} =0.
\end{array}
\end{equation}
\begin{equation}
\begin{array}{ccc}
hB^{00}&
=\frac{e^{-(2k+\Omega)}}{8\pi}\big[(2k^{\prime}+\Omega^{\prime})R^{\prime}-R^{\prime\prime}\big],
\\
hB^{01}
&=\frac{e^{-(2k+\Omega)}}{8\pi}\big[\dot{R}^{\prime}-(2\dot{k}+\dot{\Omega})R^{\prime}\big],
\\
 hB^{02}& = hB^{03}=0.
\end{array}
\end{equation}
\begin{equation}
\begin{array}{ccc}
hL^{00}& =-\frac{1}{8\pi}(RR^{\prime\prime}),\\
hL^{01}& =\frac{1}{8\pi}[R\dot{R}^{\prime}+\dot{R}R^{\prime}],\\
hL^{02}& = hL^{03}=0.
\end{array}
\end{equation}
These results agree with the results obtained in general
relativity by using these different energy-momentum complexes.
\setcounter{equation}{0}
\section{Torsion vector and axial torsion-vector}
The relation between the Weitzenb\"{o}ck connection
$\Gamma^{\sigma}_{\mu\nu}$ and the Levi-Civita connection
$\tilde{\Gamma}^{\sigma}_{\mu\nu}$ of the metric (\ref{}) is given
by
$$
\Gamma^{\sigma}_{\,\,\mu\nu}=\tilde{\Gamma}^{\sigma}_{\,\,\mu\nu}
+K^{\sigma}_{\,\,\mu\nu},
$$
where
$$
\tilde{\Gamma}^{\sigma}_{\,\,\mu\nu}=\frac{1}{2}g^{\sigma\rho}\big[
\partial_{\mu}g_{\rho\nu}+\partial_{\nu}g_{\rho\mu}-\partial_{\rho}g_{\mu\nu}\big]
$$
and
$$
K^{\sigma}_{\,\,\mu\nu} =\frac{1}{2}\big[
T^{\,\,\sigma}_{\mu\,\,\,\,\nu} +
T^{\,\,\sigma}_{\nu\,\,\,\,\mu}-T^{\sigma}_{\,\,\mu\nu}\big]
$$
is the connection tensor with $T^{\sigma}_{\,\,\mu\nu}$ given by
(\ref{tor}).\\
The torsion tensor can be decomposed into three irreducible parts
under the group of global Lorentz transformations \cite{HS}. They
are the tensor part
$$
t_{\lambda\mu\nu}=\frac{1}{2}\big(T_{\lambda\mu\nu}+T_{\mu\lambda\nu}\big)
+\frac{1}{6}\big(g_{\nu\lambda}V_{\mu}+g_{\mu\nu}V_{\lambda}\big)
$$
$$
-\frac{1}{3}g_{\lambda\mu}V_{\nu},
$$
the vector part
\begin{equation}\label{V}
V_{\mu}=T^{\nu}_{\,\,\nu\mu},
\end{equation}
and the axial-vector part
\begin{equation} \label{ax}
A^{\mu}=h_{a}^{\,\,\mu}A^{a}=\frac{1}{6}\varepsilon^{\mu\nu\rho\sigma}T_{\nu\rho\sigma}.
\end{equation}
Here the completely antisymmetric tensors
$\varepsilon^{\mu\nu\rho\sigma}$ and
$\varepsilon_{\mu\nu\rho\sigma}$ with respect to the coordinates
basis are defined by \cite{Moller}
$$
\varepsilon^{\mu\nu\rho\sigma}=\frac{1}{\sqrt{-g}}\delta^{\mu\nu\rho\sigma},
$$
$$
\varepsilon_{\mu\nu\rho\sigma}=\sqrt{-g}\delta_{\mu\nu\rho\sigma},
$$
where $\delta^{\mu\nu\rho\sigma}$ and $\delta_{\mu\nu\rho\sigma}$
are the completely antisymmetric tensor densities of weight $-1$
and $+1$, respectively, with normalization $\delta^{0123}=+1$ and
$\delta_{0123}=-1$.\\
Using (\ref{2-4}) and (\ref{torsion}) in equations (\ref{V}) and
(\ref{ax}), we get the non-vanishing components of the torsion
vector  in the following
\begin{equation}
\begin{array}{ccc}
V_{0} & = -\frac{1}{2}\big(2\dot{k}+\dot{\Omega}\big)-\frac{\dot{R}}{R},\\
V_{1} &
=-\frac{1}{2}\big(2k^\prime+\Omega^\prime\big)-\frac{R^\prime}{R},
\end{array}
\end{equation}
and axial-torsion vector
\begin{equation}
A^{\mu} = 0.
\end{equation}

\section{Discussion}
The traditional manner in which physicists have identified a total
density including the contribution from gravity has been that of
re-costing the covariant conservation laws,
$T^{\mu\nu}_{\,\,\,\,;\nu}$, into the form of an ordinary
vanishing divergence. Einstein himself changed the energy and
momentum conservation law to
$$
\frac{\partial}{\partial
x^\mu}\Big(\sqrt{-g}(T^{\mu}_{\nu}+t^{\mu}_{\nu})\Big)=0.
$$
The pseudotensor, $t^{\mu}_{\nu}$, can be changed at will and
hence there is an important ambiguity introduced regarding the
value of energy and momentum densities. In fact, the question of
the meaningfulness of energy localization in general relativity is
raised. Misner et al. \cite{MTW} argued that the energy is
localizable only for spherical systems. Cooperstock and Sarracino
\cite{CS} contradicted their viewpoint and argued that if the
energy is localizable in spherical systems then it is also
localizable for all systems. Bondi \cite{Bo} expressed that a
non-localizable form of energy is inadmissible in relativity and
its location can in principle be found. In a series of papers,
Cooperstock \cite{COO} hypothesized that in a curved space-time
energy and momentum are confined  to the region of non-vanishing
energy-momentum tensor $T^{a}_{b}$  and consequently the
gravitational waves are not carriers of energy and momentum in
vacuum space-times. This hypothesis has neither been proved nor
disproved. There are many results support this hypothesis (see for
example, \cite{Xulu}-\cite{Gad2}).
\par
These difficulties are related to the lack of precise definition
of a pseudotensor in the general relativity. As we mentioned in
the introduction many physicists have introduced different
definitions to solve this problem, but till now the problem stills
unsolved.\\
One of the approaches to solve this problem, in the context of
general relativity, is the quasilocal idea, which has been
developed by Chang, Nester and Chen \cite{CN}. According to this
idea, for each gravitational energy-momentum pseudotensor, there
is a hamiltonian boundary term. The energy-momentum defined by
such a pseudotensor does not really depend on the local value of
the reference frame, but only on the value of the reference frame
on the boundary of a region, then its quasilocal character. This
idea validates the pseudotensor approach to the gravitational
energy-momentum problem.
\par
The recent attempt to solve this problem is to replace the theory
of general relativity by another theory, concentrated on the gauge
theories for the translation group, the so called teleparallel
equivalent of general relativity.
\par
In this work we have explicitly evaluated the energy and momentum
densities in an inhomogeneous cosmological solutions of the
Einstein field equations. These solutions have an irrotational
perfect fluid, which equation of state is $p=\rho$, as source. The
energy and momentum densities are obtained in both the theory of
general relativity and the theory of teleparallel gravity. We used
different energy-momentum complexes, specifically these are the
energy-momentum complexes of Einstein, Bergmann-Thomson and
Landau-Lifshitz. We found first, in general relativity, that these
definitions do not provide the same energy and momentum
densities.\\
We were hoping that the theory of teleparallel gravity would solve
this problem. Unfortunately, using the aforementioned
energy-momentum complexes, in this theory these definitions do not
provide also the same results for the energy and momentum
densities.
\par
In the context of the theory of teleparallel gravity, the
components torsion vector and axial-vector torsion are obtained.
For the space-time under consideration the axial-vector torsion
vanishes identically.

\end{document}